\documentclass[aps,prl,superscriptaddress,reprint,groupedaddress,amsmath,amssymb,showpacs,floatfix]{revtex4-1}

\usepackage{graphicx}
\usepackage{color,soul}
\usepackage{filecontents, setspace, ctable}
\usepackage{nicefrac}

\usepackage{silence}
\newcommand\ket[1]{\ensuremath{\left|#1\right\rangle}}

\newcommand{\er}{\ensuremath{E_{\mathrm{r}}}}

\newcommand{\ttf}{\ensuremath{T/T_{F}}}

\newcommand{\eb}{\ensuremath{\epsilon_{B}}}
\newcommand{\teb}{\ensuremath{t/\epsilon_{B}}}

\newcommand{\lif} {\ensuremath{^{6}}Li}

\newcommand{\rup}{\ensuremath{R_{\uparrow}}}
\newcommand{\rdiff}{\ensuremath{R_{d}}}
\newcommand{\rdown}{\ensuremath{R_{\downarrow}}}
\newcommand{\rcore}{\ensuremath{R_{c}}}

\newcommand{\sfpp}{SF\ensuremath{_\mathrm{P}}}

\newcommand{\sfb}{SF\ensuremath{_\mathrm{0}}}
\newcommand{\npp}{N\ensuremath{_\mathrm{PP}}}
\newcommand{\nfp}{N\ensuremath{_\mathrm{FP}}}
\newcommand{\ptube}{\ensuremath{P_{t}}}
\newcommand{\po}{\ensuremath{p_{0}}}

\newcommand{\pcthree}{\ensuremath{P_{c}^{\mathrm{3D}}}}

\newcommand{\tone}{\ensuremath{\tilde{t}_{\mathrm{1D}}}}
\newcommand{\tthree}{\ensuremath{\tilde{t}_{\mathrm{3D}}}}
\newcommand{\rbar}{\ensuremath{\bar{R}}}

\newcolumntype{L}[1]{>{\raggedright\let\newline\\\arraybackslash\hspace{0pt}}m{#1}}
\newcolumntype{C}[1]{>{\centering\let\newline\\\arraybackslash\hspace{0pt}}m{#1}}

\definecolor{userblue}{rgb}{0.16,0.16,0.75}
\definecolor{userred}{rgb}{0.86,0.18,0.18}
\definecolor{usergreen}{rgb}{0,0.5,0}

\begin{document}


\title{1D to 3D Crossover of a Spin-Imbalanced Fermi Gas}
\author{Melissa C. Revelle}
\author{Jacob A. Fry}
\author{Ben A. Olsen}
\altaffiliation{Current address: Department of Physics, University of Toronto, Ontario M5S 1A7, Canada}
\author{Randall G. Hulet}
\affiliation{Department of Physics \& Astronomy and Rice Center for Quantum Materials, Rice University, Houston, TX 77005, USA}

\date{November 1, 2016}

\begin{abstract}
We have characterized the one-dimensional (1D) to three-dimensional (3D) crossover of a two-component spin-imbalanced Fermi gas of \lif{} atoms in a 2D optical lattice by varying the lattice tunneling and the interactions.
The gas phase separates, and we detect the phase boundaries using \textit{in situ} imaging of the inhomogeneous density profiles.
The locations of the phases are inverted in 1D as compared to 3D, thus providing a clear signature of the crossover.
By scaling the tunneling rate $t$ with respect to the pair binding energy \eb{}, we observe a collapse of the data to a universal crossover point at a scaled tunneling value of $\tilde{t}_{c} = 0.025(7)$.
\end{abstract}

                              
\maketitle


Atomic Fermi gases prepared in two hyperfine sublevels realize a quasi-spin-\nicefrac{1}{2} system, for which the two states may be denoted as \ket{\uparrow} and \ket{\downarrow}.  
Spin-imbalanced Fermi gases, where the number of spin-up atoms, $N_{\uparrow}$, exceeds the number of spin-down atoms, $N_{\downarrow}$, have been studied extensively in recent years, largely motivated by a search for exotic superfluid phases~\cite{Giorgini08, Radzihovsky10, Bloch08}. 
One such superfluid, the Fulde-Ferrell-Larkin-Ovchinnikov (FFLO) phase~\cite{Fulde64, Larkin64}, has not been conclusively observed in three dimensions (3D) but is believed to occupy a large portion of the one-dimensional (1D) phase diagram~\cite{Orso07, Hu07}.
Measurements have confirmed that the 1D phase diagram is consistent with theories exhibiting FFLO~\cite{Liao10}, but direct evidence for this phase remains elusive. 
Since the FFLO phase is expected to be more robust to quantum and thermal fluctuations in higher dimensions, attention has focused on the dimensional crossover~\cite{Parish07, Zhao08, Sun13, Heikkinen14}.

\begin{figure}[ht]
\centering
\includegraphics[width=\columnwidth]{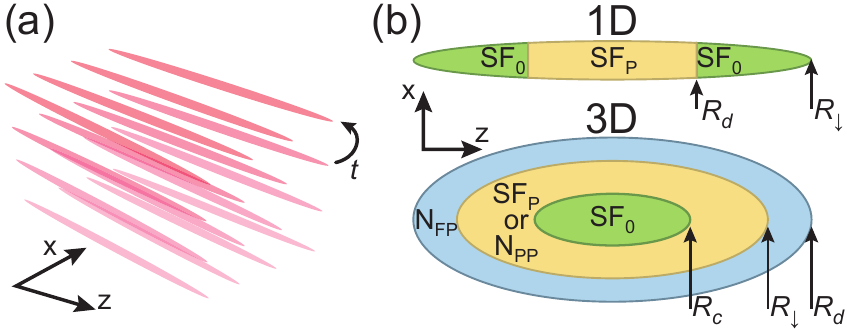}
\caption{(Color online) 
(a) Schematic of an array of 1D coupled tubes formed by a 2D optical lattice. 
The tunneling rate $t$ between the tubes increases with decreasing optical lattice depth.
(b) Schematic of phase separation for a trapped spin-imbalanced Fermi gas in 1D (top) and in 3D (bottom) at zero temperature. 
In 1D, the central region is an FFLO partially-polarized superfluid (\sfpp{}), with balanced superfluid (\sfb{}) wings for small polarization $P$.
In 3D, for $P < \pcthree{}$, a central \sfb{} core is surrounded by an \sfpp{} or normal partially-polarized (\npp{}) phase depending on interactions, and finally an \nfp{} outer shell. 
The arrows indicate phase boundaries. }
\label{fig:cartoon}
\end{figure}

A crossover between 1D and 3D regimes may be realized by simply varying the confinement aspect ratio~\cite{Castellani94, Gorlitz01, Astrakharchik02, Gerbier04, Armijo11}.
A complementary dimensional crossover occurs by varying the tunneling between tubes aligned in an array, as depicted in Fig.~\ref{fig:cartoon}(a).
Such a geometry, which may be achieved using ultracold atoms in an optical lattice, is more analogous to some material systems, such as carbon nanotube bundles~\cite{Calbi03} and spin-$\nicefrac{1}{2}$ magnet chains~\cite{Carlson00, Pan14}. 
The bundle will cross over from an array of independent 1D tubes for small tunneling $t$, to a 3D system as $t$ is increased~\cite{Mortiz05, Vogler14}.
We have employed this geometry to determine the crossover value of $t$ for a spin-imbalanced Fermi gas with various interaction strengths and find a striking universality in the crossover location. 


Trapped Fermi gases with spin-imbalance have been observed to phase separate at low temperatures in both 3D~\cite{Partridge06, Shin06, Shin08PhaseDiag, Navon10, Olsen15} and in 1D~\cite{Liao10}, although in a qualitatively different manner.  
As shown in Fig.~\ref{fig:cartoon}(b), phase separation in 1D results in a partially-polarized superfluid (\sfpp{}) central core with wings that are either a fully-paired superfluid (\sfb{}) or a fully-polarized (\nfp{}) phase, depending on the spin-polarization $P$ in the tube.  
Theory indicates the \sfpp{} phase is an FFLO superfluid~\cite{Orso07, Hu07}.
It was previously shown that the axial radii of the minority state distribution, \rdown{}, and the spin-difference distribution, \rdiff{}, determine the 1D phase boundaries~\cite{Liao10}, as indicated in Fig.~\ref{fig:cartoon}(b).  
\rdiff{} corresponds to the boundary between the \sfpp{} core and the \sfb{} wings since the spin-difference density is zero in the \sfb{} wings. 
\rdiff{} goes to zero for $P = 0$, but moves to larger axial radius with increasing $P$ until the polarized core encompasses the entire cloud. 
At this polarization, the entire tube is in the \sfpp{} phase and $\rdiff{} = \rup{} =\rdown{}$, where \rup{} is the axial radius of the majority state distribution~\cite{Orso07, Liao10}. 
At even larger $P$, the boundary between the \sfpp{} core and the \nfp{} wings is defined by \rdown{}. 

Phase separation in a trapped 3D gas at low temperature results in a shell structure, also depicted in Fig.~\ref{fig:cartoon}(b).  
The relative location of the phases in 3D is largely inverted compared to 1D.  
The center of the cloud in 3D is a balanced \sfb{} phase for $P$ less than a critical polarization \pcthree{}, beyond which superfluidity is suppressed~\cite{Clogston62, Chandrasekhar62, Zwierlein06, Shin06, Shin08PhaseDiag, Bertaina09, Navon10, Olsen15}.
In addition to being spin-balanced, the previous observation of quantized vortices proved that the core was superfluid~\cite{Zwierlein06}.  
The boundary between the unpolarized \sfb{} phase and a polarized \sfpp{}, or a partially-polarized \npp{} normal phase (depending on interactions), is defined by the axial “core” radius \rcore{} where the spin-difference density first rises above zero from the center of the cloud~\cite{Shin08PhaseDiag, Bertaina09, Olsen15}.  
A fully polarized normal shell (\nfp{}) sits outside the partially-polarized region and the boundary between them is given by \rdown{}. 
The outer boundary of the cloud, going to vacuum, is defined by $\rup{} = \rdiff{}$.

The distinction between phase separation in 1D and 3D can be used to signal the dimensionality of the system. 
By varying tube coupling and interactions the location of the dimensional crossover will be revealed by the central polarization at small $P$:  a partially polarized core is 1D-like, while the presence of an unpolarized core at small $P$ is 3D-like~\cite{Footnote01}.



As described in detail previously~\cite{Partridge06, Liao10}, our experiment employs the lowest two hyperfine sublevels of~\lif{}, the \ket{F = \nicefrac{1}{2}, \, m_F = \nicefrac{1}{2}} state, designated as \ket{\uparrow}, and  the \ket{F = \nicefrac{1}{2}, \, m_F = - \nicefrac{1}{2}} state, designated as \ket{\downarrow}.
These correspond to the majority and the minority states, respectively.
The atoms are prepared in a population imbalanced mixture and evaporatively cooled in an optical trap~\cite{Liao10}. 
A 2D optical lattice is formed by an orthogonal pair of retro-reflected laser beams at a wavelength $\lambda$ of $1064\,$nm.
The lattice depth $V_{L}$ may be controlled up to a maximum value of 12\,\er{} using liquid crystal retarders (LCRs) to rotate the polarization of the retro-reflected beams with respect to the incoming beams. 
Here, $\er{} =\hbar^{2}k^{2}/2m$ is the lattice recoil energy, $k=2\pi/\lambda$, and $m$ is the atomic mass.
The axial ($z$) potential is approximately harmonic with a frequency $\omega_{z}$ that varies linearly with $V_{L}$ from $(2 \pi)197$\,Hz for $V_{L}=2.5$\,\er{} to $(2 \pi )256\,$Hz for $V_{L}=12$\,\er{}. 
We find that the mean number of \ket{\uparrow} atoms in the central tube, $N_{\uparrow}$,  is between 160 and 240 for small ($<5\%$) polarizations, but it decreases for larger polarizations due to inefficient evaporation.
The interaction strength between the two states is tuned via the wide Feshbach resonance located at $B = 832.2\,$G~\cite{Houbiers98, Zurn13}.
We independently control both $t$ and the atomic interactions by varying $V_{L}$ and the magnetic field, $B$.

The criteria for each tube to be in the 1D regime are that both the Fermi energy $E_{F} = k_{B} T_{F} = N_{\uparrow} \hbar \omega_{z}$ and the temperature $T$ be small compared to the transverse confinement energy: $E_{F},k_{B}T \ll \hbar \omega_{\perp}$, where $\omega_{\perp}$ is the transverse frequency within a tube.
Additionally, when $t \ll T,E_{F}$ the entire bundle behaves as an array of individual 1D tubes~\cite{Liao10}.
The value of $E_{F}/\hbar \omega_{\perp}$ in the central tube of our experiment is between $0.2$ and $0.4$.
We measure $\ttf{} = 0.05$ before transferring the atoms into the lattice by fitting the \textit{in situ} column density profiles to finite temperature Thomas-Fermi distributions.
The entropy in the lattice may be bounded by this measurement and by measuring the temperature in the trap after ramping the lattice on and back off with the LCRs.
We measure a maximum temperature of $\ttf{} = 0.16$ after this round-trip, which is consistent with our previous 1D experiment~\cite{Liao10}. 

\begin{figure}[ht]
\centering
\includegraphics[width=\columnwidth]{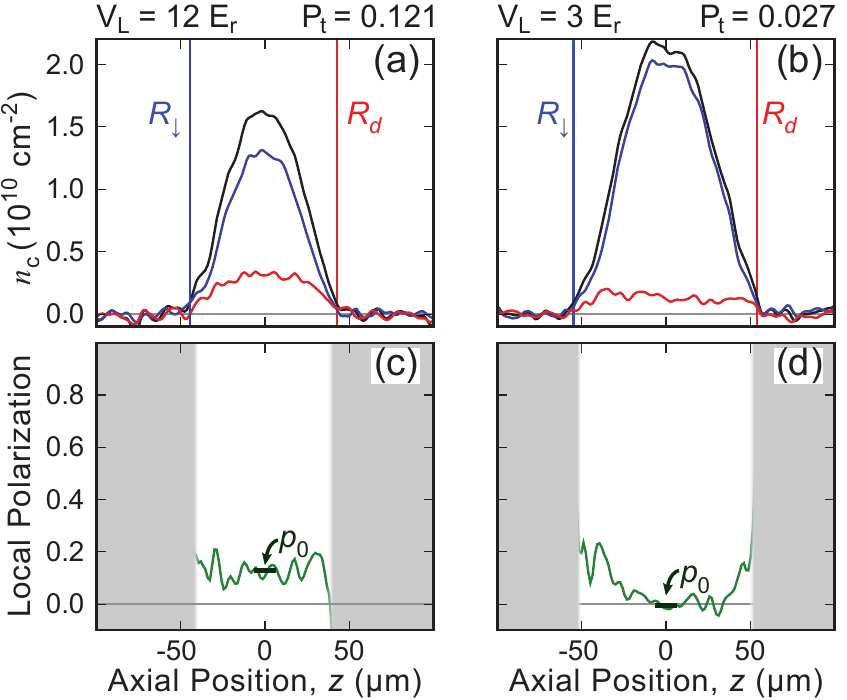}
\caption{(Color online) (a, b) Column density profiles $n_{c}(0,0,z)$ of spin-imbalanced gases.
The $n_{c}$ are smoothed in $x$ and $z$ using a Gaussian function with a width of $5.3\,\mu$m before taking a cut along the $z$-axis. 
Both data sets were taken at $B = 890\,$G, corresponding to $a_{\mathrm{3D}} = -8610\,a_{0}$.
The scaled tunneling (defined in text) is $\teb{} = 0.004$ for the first column and $0.065$ for the second. 
The \ket{\uparrow}, \ket{\downarrow}, and the difference distributions are indicated by the black, blue, and red curves, respectively. 
The radii are extracted on both sides of the cloud by finding the radius at which a phenomenological fit to the $n_{c}$ rises by one standard deviation above the mean background level.  
The radii extracted from each side are averaged together.
(c, d) The corresponding local polarization $p(0,0,z)$ profiles are found using a weighted average of the central 18 tubes.
\po{} is the average of the central $13\,\mu$m region along $z$.
$N_{\downarrow}$ is consistent with the background noise in the gray region and thus, the local polarization is poorly defined there. 
The entire cloud in (a) and (c) is \sfpp{}, and $\rdown{} \simeq \rdiff{}$ as a consequence, while in (b, d), there is an extended region of \sfb{} in the center of the cloud ($\po{}=0$), then a partially-polarized region, \sfpp{} or \npp{}.
$\rdown{} \approx \rdiff{}$ in this 3D-like example since \ptube{} is small.}
\label{fig:profiles2}
\end{figure}

We use \textit{in situ} phase-contrast-polarization imaging~\cite{Bradley97} to measure the column density distributions $n_{c}(x,z)$ for each spin state by two successive probe pulses, each of different near-resonant detuning from the $^{2}P_{\nicefrac{3}{2}}$ excited state ~\cite{Liao10}. 
The probe pulse duration is ${\sim} 5 \,\mu$s and the time between the two pulses is ${\sim} 1 \,\mu$s. 
The probe beams propagate along the $y$-axis, perpendicular to the tubes which are aligned along the $z$-axis.
We use an inverse Abel transform to obtain the full density distribution of the cloud, $n(x,y,z)$, from the $n_{c}(x,z)$ by making use of the quasi-cylindrical symmetry about the $z$-axis.
The number of atoms per spin state in the central tube, $N_{\uparrow}$ and $N_{\downarrow}$, are extracted from the densities and are used to calculate the central tube polarization $\ptube{}=(N_{\uparrow}-N_{\downarrow})/(N_{\uparrow}+N_{\downarrow})$.
Figures~\ref{fig:profiles2}(a) and (b) show axial ($z$) cuts of \textit{in situ} column density images for two different lattice depths for both spin states and for the spin-difference.

The radii \rdown{} and \rdiff{} may be extracted from the $n(x,y,z)$ or obtained directly from the $n_{c}(x,z)$ distributions by assuming the validity of the local density approximation (LDA) in the radial direction.
Since the chemical potential of each spin state is largest for the central tube, the phase boundaries, \rdown{} and \rdiff{}, are largest for the central tube and decrease radially. 
We therefore use the central axial cut ($x=0$) of the $n_{c}(x,z)$ to locate \rdown{} and \rdiff{} corresponding to the central tube. 
These are indicated in Figs.~\ref{fig:profiles2}(a) and (b).


Figure~\ref{fig:profiles2}(a) shows a 1D-like profile, where the spin-difference column density profile is approximately parabolic, in contrast to Fig.~\ref{fig:profiles2}(b) which is consistent with 3D phase separation.
The distinction between 3D and 1D phase separation is confirmed by examination of the local polarization $p(0,0,z) = (n_{\uparrow}(0,0,z)-n_{\downarrow}(0,0,z))/(n_{\uparrow}(0,0,z)+n_{\downarrow}(0,0,z))$, where $n_{\uparrow}$ and $n_{\downarrow}$ are the densities of each state obtained from the inverse Abel transformed data.
The polarization at the center, $\po{}=p(0,0,0)$, reveals the central phase.
In Fig.~\ref{fig:profiles2}(c), $\po{} > 0$, corresponding to a partially-polarized central phase consistent with 1D phase separation, while Fig.~\ref{fig:profiles2}(d) shows an example with $\po{}=0$, and is therefore consistent with 3D-like phase separation containing a \sfb{} core.

\begin{figure}[ht]
\centering
\includegraphics[width=\columnwidth]{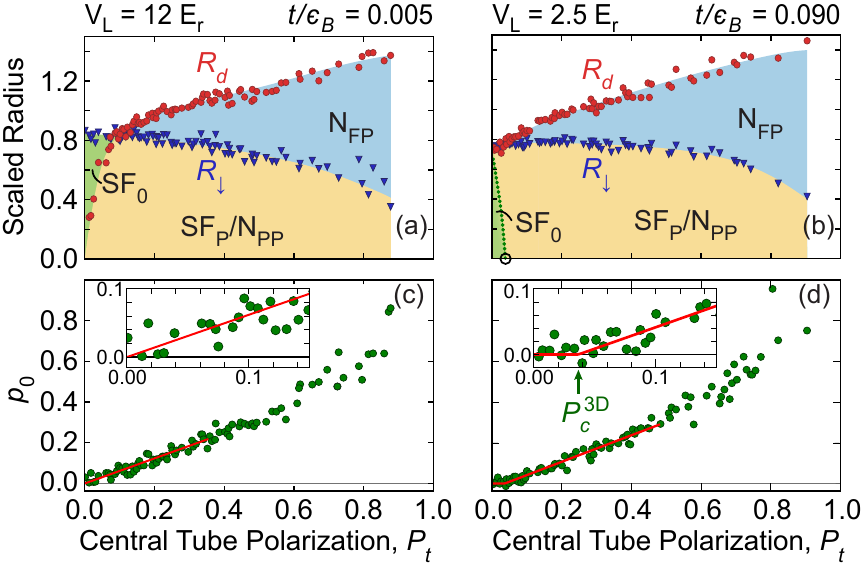}
\caption{(Color online) (a) 1D- and (b) 3D-like phase diagrams for $B = 940\,$G. 
\rdown{} (\textcolor{userblue}{\scriptsize$ \blacktriangledown$}) and \rdiff{} ($\textcolor{userred}{\bullet}$) are scaled by $N^{1/2}l_{z}$~\cite{Orso07, Liao10}, where $l_{z}=\sqrt{\hbar/m \omega_{z}}$ is the axial harmonic oscillator length and $N=N_{\uparrow} + N_{\downarrow}$.
The colored regions correspond to the indicated phases.
In (b), the open circle indicates the measured \pcthree{} from (d).
The dotted line is an extrapolation from \pcthree{}.
(c, d) The local central polarization \po{} vs. \ptube{}, used to find \pcthree{}.
The insets show the central region near \pcthree{}.
The solid red line is a fit to the data to find \pcthree{}, using a function with a bilinear slope~\cite{Olsen15}.
The green vertical arrow indicates \pcthree{}. 
Each data point is the average of ${\sim}10$ experimental realizations, binned with width $\Delta \ptube{}=0.005$.}
\label{fig:diag_combined}
\end{figure}

Two examples of phase diagrams constructed from the radii \rdiff{} and \rdown{} are presented in Figs.~\ref{fig:diag_combined}(a) and (b). 
Figure~\ref{fig:diag_combined}(a) corresponds to a relatively deep lattice, with $V_{L} = 12 \,\er{}$, that exhibits a 1D-like phase diagram with a partially-polarized core, similar to those reported in Ref~\cite{Liao10}. 
The distinguishing characteristics of the 1D-like phase diagram are 1) \rdiff{} goes to zero as \ptube{} goes to zero, and 2) \rdiff{} crosses \rdown{} at a non-zero \ptube{}. 
Figure~\ref{fig:diag_combined}(b) shows an example of a 3D-like phase diagram where the centrally located phase at small \ptube{} is \sfb{}, and \rdiff{} decreases with decreasing \ptube{} until meeting \rdown{} at small $\ptube{}$.

\begin{figure*}[t]
\centering
\includegraphics[width=\textwidth]{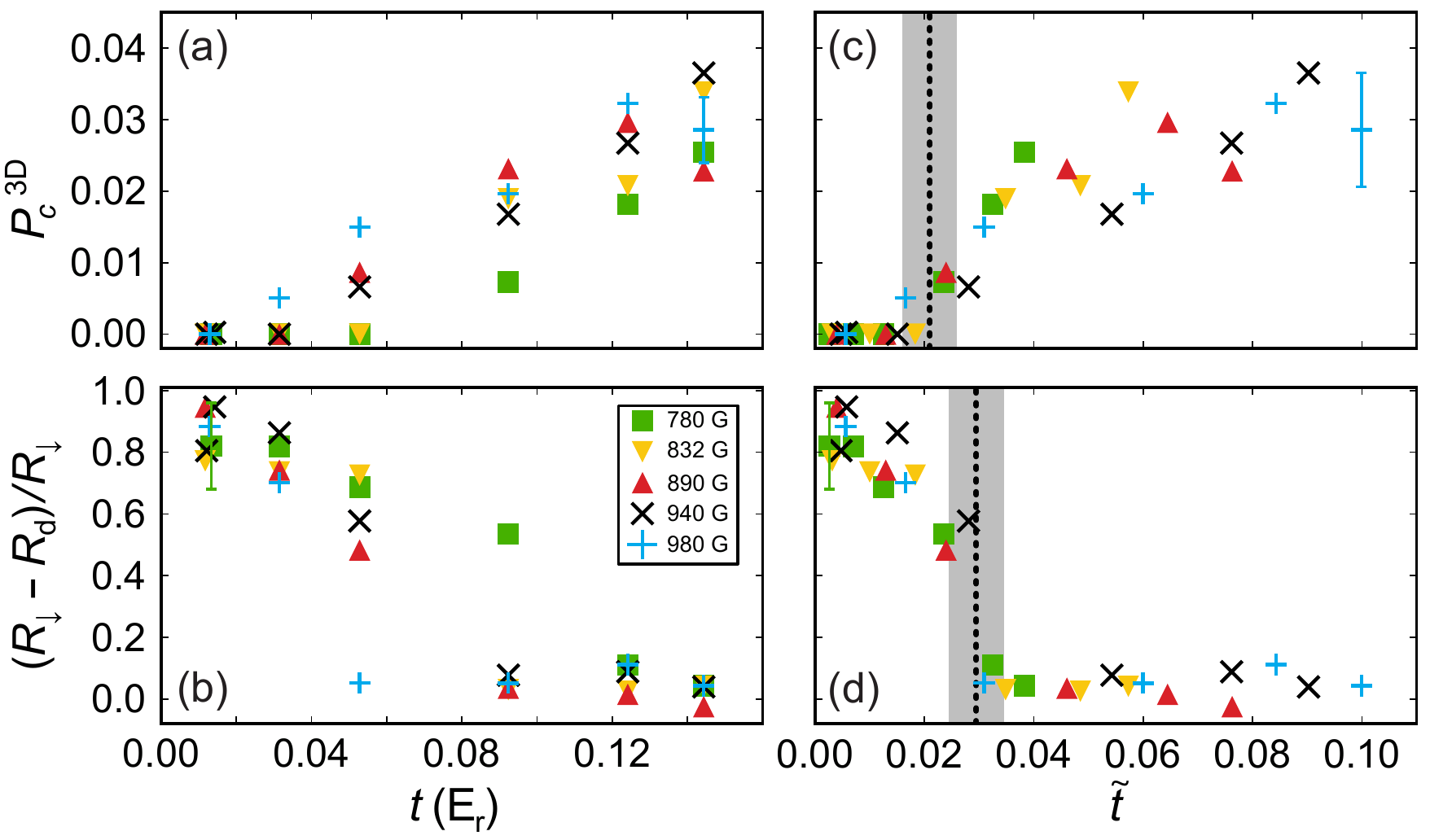}
\caption{(Color online) (a) \pcthree{} and (b)  \rbar{} vs. $t$.
Ordered from lowest to highest field, the corresponding $a_{\mathrm{3D}}$ are: $6170\,a_{0}$, unitarity, $-8610\,a_{0}$, $-5360\,a_{0}$, and $-4340\,a_{0}$, in units of the Bohr radius $a_{0}$. 
The corresponding ranges of \eb{}, depending on lattice strength, are: $3.8-5.2\,\er{}$, $2.5-3.7\,\er{}$, $1.9-2.9\,\er{}$, $1.6-2.5\,\er{}$, and $1.4-2.3\,\er{}$, respectively.
(c) \pcthree{} and (d) \rbar{} vs. the scaled tunneling rate $\tilde{t} = \teb{}$, showing data collapse.
The dotted line in (c) indicates $\tthree{} =0.021(5)$, the value above which the gas has an \sfb{} core.
The suppression of 1D behavior occurs at  $\tone{} =0.029(5)$, indicated by the dotted line in (d).
The gray band indicates the uncertainty range in locating \tthree{} and \tone{}. 
These uncertainties result from the indicated vertical error bars (a few representative examples are shown) which arise from the fits, as well as systematic uncertainty in \ptube{} which is estimated from the standard error of the mean of 10 images known to be balanced. }
\label{fig:PCall}
\end{figure*}


We identify phase separation in 3D by the presence of a superfluid core that is suppressed above a critical polarization \pcthree{}~\cite{Olsen15, Shin06}.
\pcthree{} is defined to be the \ptube{}, above which, \po{} begins to rise from zero.
For $\pcthree = 0$, there is no balanced core for any \ptube{}, and thus the gas is 1D-like.
Figure~\ref{fig:diag_combined}(c) shows \po{} corresponding to the 1D phase diagram of Fig.~\ref{fig:diag_combined}(a), where \po{} increases linearly with \ptube{}.
A crossover to 3D occurs when $V_{L}$ is decreased so that $t$ becomes sufficiently large to produce a kink in \po{} vs. \ptube{}, as seen in Fig.~\ref{fig:diag_combined}(d).
The open circle in Fig.~\ref{fig:diag_combined}(b) indicates the measured \pcthree{} from Fig.~\ref{fig:diag_combined}(d).

Figure~\ref{fig:PCall}(a) shows \pcthree{} vs. $t$ for several interaction strengths. 
We calculate $t$ from the eigenenergies of the 1D Hamiltonian~\cite{Morsch06}.
The calculated single particle tunneling rate includes nearest neighbor and next-nearest neighbor contributions, where the latter becomes significant at lattice depths below 5\,\er{}.
Comparing \rdiff{} and \rdown{} as \ptube{} goes to zero is also an indicator of dimensionality. 
The normalized ratio $\rbar{} = (\rdown-\rdiff)/\rdown$ goes to 1 in 1D as \rdiff{} goes to 0, but in 3D, \rbar{} goes to 0 as \rdiff{} approaches \rdown{}. 
In Fig.~\ref{fig:PCall}(b), we plot \rbar{} vs. $t$ for the same interaction strengths. 
Figures~\ref{fig:PCall}(a) and (b) show that the 3D regime is attained for large $t$, as expected, but also for larger $B$, corresponding to weaker attractive interactions and thus larger chemical potentials. 
We believe that the interaction dependence arises from the suppression of pair tunneling in the BEC regime (smaller $B$) where \eb{} is large, thus making the BEC regime more 1D-like~\cite{Sun13}. 

In Figures~\ref{fig:PCall}(c) and (d), we replot the data against the scaled tunneling rate $\tilde{t} = \teb{}$, where \eb{} is the pair binding energy calculated from~\cite{Bergeman03}:
\begin{equation}
\frac{\sqrt{2} l_{\perp}}{a_{\mathrm{3D}}}= -\zeta\left[\frac{1}{2},\frac{-\eb{}}{2 \hbar \omega_\perp} \right],
\end{equation}
where $\zeta$ is the Hurwitz zeta function. 
This solution depends on the transverse harmonic oscillator length $l_\perp=\sqrt{\hbar/m \omega_{\perp}}$, as well as the 3D $s$-wave scattering length $a_{\mathrm{3D}}$. 
When scaled in this way, the data collapse onto a single curve, thus demonstrating the universality of the crossover~\cite{Parish07}.
As shown in Fig.~\ref{fig:PCall}(c), the suppression of the \sfb{} core occurs at $\tthree{} = 0.021(5)$. 
The uncertainty is a combination of the error from fitting \pcthree{} and the systematic uncertainty in measuring \ptube{}. 
We used only small \ptube{} ($< 25\%$) to determine \pcthree{} in order to justify the assumption of a linear dependence of \po{} on \ptube{}. 
The data for \rbar{} also collapse to a single curve when plotted vs.~$\tilde{t}$, as shown in Fig.~\ref{fig:PCall}(d).  
We find that \rbar{} decreases sharply at $\tone{}= 0.029(5)$, as the gas transitions from 1D to 3D.
Although \tone{} and \tthree{} may be distinct, the difference between them is within their mutual uncertainties, so we combine our two measurements of the crossover location to give $\tilde{t}_c = 0.025(7)$.

A mean field analysis has predicted that the phase boundary between the \sfb{} core and the \nfp{} phase corresponds to a first order transition~\cite{Parish07}. 
Due to noise in the inverse Abel transformed data, however, we are unable to directly observe a jump in the local polarization.
This could also be a consequence of finite $T$. 
Mean-field theory also predicts that the 3D to 1D crossover may be driven by increasing the chemical potential $\mu$~\cite{Parish07}.  
The slope of this boundary, however, is very steep in the $\mu$ vs.~$h$ plane, where $h$ is the chemical potential difference, thus causing the location of this transition to be at very large $\mu$.
Since our measurements are performed in the regime where $\ptube{} \rightarrow 0$, or equivalently $h \rightarrow 0$, a transition back to 1D could only occur at such a large $\mu$ that the 1D criterion for each tube would not hold.
Our experiment finds the location of the dimensional crossover $\tilde{t}_{c}$ at the center of the trap, where the total variation in the measured densities is no more than a factor of 1.6 for all of the data. 
$\tilde{t}_{c}$ should depend on density, but we have not measured this dependence.


In conclusion, our results show that the 1D to 3D crossover occurs at a universal value of the scaled tunneling, $\tilde{t}_{c}$. 
Looking towards the future, the crossover region is predicted to be the most robust against fluctuations in FFLO wavenumber and temperature~\cite{Parish07}, suggesting the most fruitful parameter region to search for the FFLO phase is the quasi-1D regime near $\tilde{t}_{c}$.

\begin{acknowledgments}
The authors would like to thank Erich Mueller, Dan Sheehy, David Huse, and Meera Parish for many valuable discussions. This work was supported by grants from the NSF (Grant PHY-1607215), the Welch Foundation (Grant No. C-1133), an ARO-MURI (Grant No. W911NF-14-1-0003), and the ONR.
\end{acknowledgments}


\begin{thebibliography}{37}
\bibitem [{Foo()}]{Footnote01}%
  \BibitemOpen
  \href@noop {} {}\bibinfo {note} {The \sfb{} core disappears in the deep BEC
  regime in 3D where at sufficiently large $P$ the gas becomes a
  partially-polarized (\sfpp{}) Bose-Fermi mixture~\cite{Bertaina09, Olsen15}.
  Here we explore the regime near the dimensional crossover where \pcthree{} is
  small}\BibitemShut {NoStop}%
\end{thebibliography}
\end{document}